\documentclass{article}

\PassOptionsToPackage{numbers, compress}{natbib}
 \usepackage[preprint]{unireps_2025}

\usepackage[utf8]{inputenc} 
\usepackage[T1]{fontenc}    
\usepackage{hyperref}       
\usepackage{url}            
\usepackage{booktabs}       
\usepackage{amsfonts}       
\usepackage{nicefrac}       
\usepackage{microtype}      
\usepackage[dvipsnames]{xcolor}
\usepackage{amsmath}
\usepackage{amssymb}
\usepackage{bbm}
\usepackage{algorithm}
\usepackage{algpseudocode}
\usepackage{graphicx}
\usepackage{url}
\usepackage{cleveref}
\usepackage{listings}
\usepackage{subcaption}  
\usepackage{adjustbox}

\hypersetup{
    colorlinks=true,
    colorlinks=NavyBlue,
    linkcolor=NavyBlue,
    filecolor=NavyBlue,      
    urlcolor=NavyBlue,
    citecolor=NavyBlue
    }

\definecolor{codebg}{HTML}{F7F7F7}
\definecolor{frame}{HTML}{DDDDDD}
\definecolor{kw}{HTML}{005CC5}
\definecolor{str}{HTML}{032F62}
\definecolor{com}{HTML}{6A737D}
\definecolor{num}{HTML}{1F6FEB}
\definecolor{ident}{HTML}{24292E}

\lstdefinestyle{sslpython}{
  language=Python,
  backgroundcolor=\color{codebg},
  basicstyle=\ttfamily\small,
  keywordstyle=\bfseries\color{kw},
  stringstyle=\color{str},
  commentstyle=\itshape\color{com},
  identifierstyle=\color{ident},
  numberstyle=\scriptsize\color{com},
  numbers=left,
  numbersep=10pt,
  frame=single,
  rulecolor=\color{frame},
  frameround=tttt,
  tabsize=2,
  showstringspaces=false,
  keepspaces=true,
  breaklines=true,
  breakatwhitespace=true,
  columns=fullflexible,
  upquote=true,
  captionpos=b
}

\title{\texttt{stable-pretraining-v1}:\\ Foundation Model Research Made Simple}

\author{%
  Randall Balestriero$^{1}$\thanks{Correspondence to \texttt{rbalestr@brown.edu}} , Hugues Van Assel$^2$, Sami BuGhanem$^1$, Lucas Maes$^3$\\
  $^1$Brown University, $^2$Genentech, $^3$Mila \& Université de Montréal\\
}

\begin{document}

\maketitle

\begin{abstract}
Foundation models and self-supervised learning (SSL) have become central to modern AI, yet research in this area remains hindered by complex codebases, redundant re-implementations, and the heavy engineering burden of scaling experiments. We present \texttt{stable-pretraining}, a modular, extensible, and performance-optimized library built on top of PyTorch, Lightning, Hugging Face, and TorchMetrics. Unlike prior toolkits focused narrowly on reproducing state-of-the-art results, \texttt{stable-pretraining} is designed for flexibility and iteration speed: it unifies essential SSL utilities—including probes, collapse detection metrics, augmentation pipelines, and extensible evaluation routines—within a coherent and reliable framework. A central design principle is \emph{logging everything}, enabling fine-grained visibility into training dynamics that makes debugging, monitoring, and reproducibility seamless. We validate the library by demonstrating its ability to generate new research insights with minimal overhead, including depth-wise representation probing and the analysis of CLIP degradation under synthetic data finetuning. By lowering barriers to entry while remaining scalable to large experiments, \texttt{stable-pretraining} aims to accelerate discovery and expand the possibilities of foundation model research. The source code is available at \url{https://github.com/rbalestr-lab/stable-pretraining}.
\end{abstract}

\begin{figure}[H]
    \centering
    \includegraphics[scale=0.56]{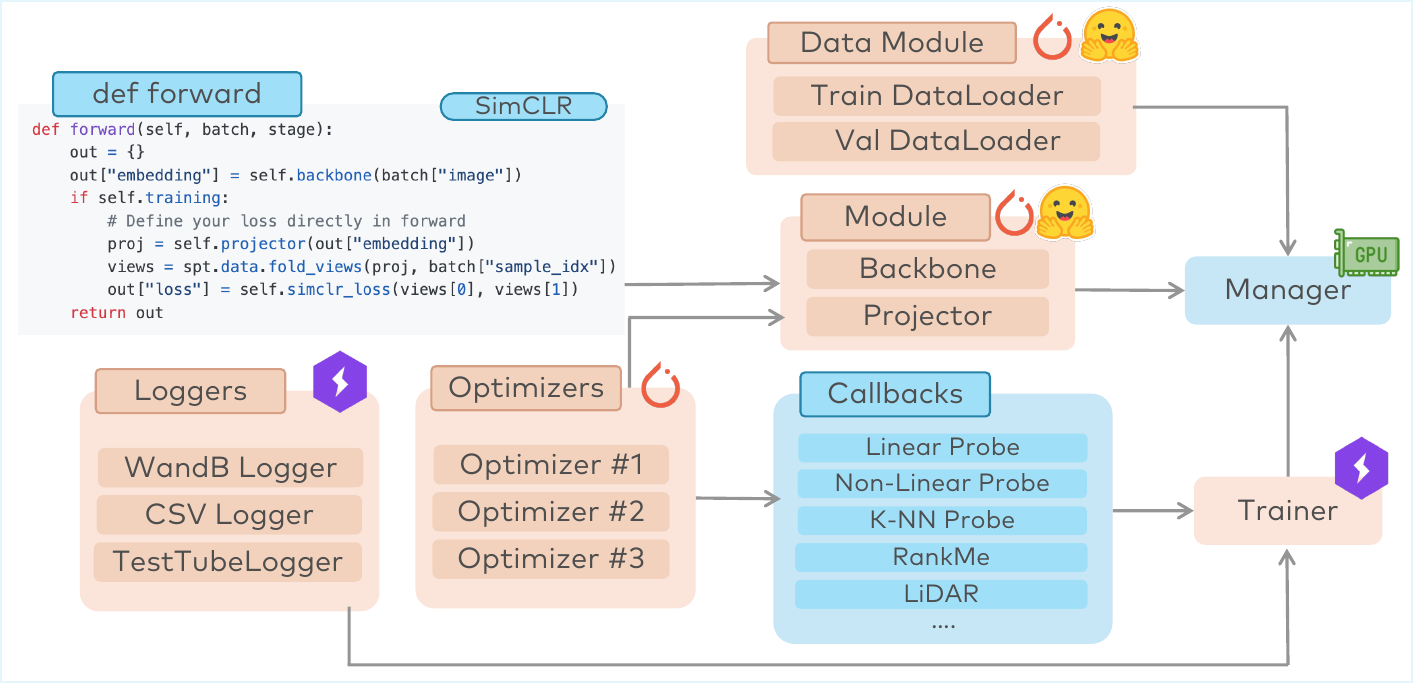}
    \caption{Overview of \texttt{stable-pretraining}.}
    \label{fig:pipeline}
\end{figure}

\section{Introduction}

Foundation models have transformed artificial intelligence in the past decade, powering breakthroughs across vision, language, and multimodal learning. Yet, despite this progress, research on foundation models remains uniquely challenging. Unlike conventional supervised learning, it requires large-scale datasets, multi-GPU training setups, and intricate monitoring of training dynamics. Researchers must navigate debugging difficulties, collapse detection, careful hyperparameter tuning, and complex evaluation protocols \citep{balestriero2023cookbook}—none of which are readily supported in mainstream frameworks like PyTorch \citep{pytorch}, Lightning\citep{lightning}, or Hugging Face\citep{wolf-etal-2020-transformers, lhoest-etal-2021-datasets}. As a result, even simple experiments often demand starting from massive, monolithic codebases such as DINOv2 \citep{oquab2023dinov2} or MAE \citep{he2022masked}. These repositories are difficult to extend, tightly coupled to specific engineering choices, and slow to prototype with—creating a bottleneck for innovation. Compounding the problem, many research groups repeatedly re-implement the same essential components: data augmentation pipelines, training loops, probes, loss functions, or evaluation metrics. This redundancy is not only inefficient but also increases the likelihood of bugs, inconsistencies, and incomparable evaluation results across the community. The consequence is {\em a research ecosystem constrained to incremental improvements, with limited room for rapid exploration of new ideas}.

Several prior libraries have attempted to address these challenges, such as VISSL~\citep{goyal2021vissl}, solo-learn~\citep{solo-learn}, or lightly~\citep{susmelj2020lightly}. However, these toolkits share important limitations, e.g., they are static by design, focusing on reproducing established methods rather than supporting new research exploration. Moreover, VISSL and solo-learn are no longer actively maintained, with their last commits dating back to 2022 and 2023, respectively. Lightly, on the other hand, separates SSL functionality from training utilities, many of which are only accessible through paid membership. Lastly, none of these frameworks treat monitoring and debugging as first-class concerns, leaving researchers to repeatedly engineer their own probes, evaluation pipelines, or collapse detection metrics. As a result, {\em existing solutions only partially reduce the engineering burden and do not fully support the rapid, exploratory workflows needed for foundation model research}.

To address these challenges, we present \texttt{stable-pretraining}, a library purpose-built for rapid and scalable foundation model research. Built on top of PyTorch, Lightning, Hugging Face, and TorchMetrics \citep{Detlefsen2022TorchMetrics}, it combines the reliability of widely adopted frameworks with specificities required for foundation model training, typically absent elsewhere. Unlike prior toolkits focused narrowly on reproducing state-of-the-art results, \texttt{stable-pretraining} is designed for flexibility and iteration speed. Its modular framework consolidates critical SSL components—including probes (linear, non-linear, $k$-NN), collapse detection metrics (RankMe \citep{garrido2023rankme}, LiDAR \citep{thilak2023lidar}), and extensible evaluation utilities—into a unified, performance-optimized system. At its core, \texttt{stable-pretraining} logs every aspect of training and evaluation, providing fine-grained monitoring and transparent feedback that facilitates debugging, reproducibility, and deeper insights from training dynamics. Our goal is to expand what is possible in foundation model research: to accelerate discovery, foster reproducibility, and empower the community to explore beyond today’s incremental progress.

\begin{table}[h!]
\centering
\caption{Linear probe top-1 accuracy across multiple datasets.}
\begin{adjustbox}{max width=\textwidth}
\begin{tabular}{lccccccccccc}
\toprule
Method & Arch. & DTD & aircraft & cars & cifar10 & cifar100 & flowers102 & food101 & galaxy10 & pets & avg. \\
\toprule
I-JEPA \citep{assran2023self} & ViT-H & 73.62 & 56.45 & 58.93 & 97.77 & 86.93 & 85.76 & 81.06 & 62.93 & 92.94 & 77.37 \\
DINO \citep{caron2021emerging} & ViT-S & 77.29 & 72.92 & 75.86 & 97.12 & 85.27 & 95.13 & 84.81 & \textbf{68.91} & 95.00 & 83.59\\
DINOv2 \citep{oquab2023dinov2} & ViT-S & \textbf{80.43} & \textbf{80.56} & \textbf{84.21} & \textbf{97.75} & \textbf{88.04} & \textbf{99.56} & \textbf{90.52} & 67.60 & \textbf{95.67} & \textbf{87.15} \\
\hline
\end{tabular}
\end{adjustbox}
\label{table:probe}
\end{table}

\section{\texttt{stable-pretraining}: An Overview}

\texttt{stable-pretraining}'s focus is to alleviate the tedious process of assembling a foundation model research codebase. We argue that the lack of such library poses an important limitation in current research as the barrier to entry has become insurmountable. With our solution, the time from research idea to first sign of success of failure is drastically reduced. In the following sections, we first outline the design choices behind \texttt{stable-pretraining}. We then highlight our research utilities by presenting two simple yet previously unverified experimental insights in self-supervised learning.

\subsection{Structure}

\Cref{fig:pipeline} provides an overview of \texttt{stable-pretraining}. Our design philosophy is simple: reuse what the community already trusts, and build only what is missing to perform efficient research. Components shown in blue represent modules we specifically developed, while those in orange are borrowed and adapted from proven third-party libraries such as Lightning, Hugging Face, and PyTorch. At the center of the pipeline is the \texttt{Manager}, a lightweight controller that works in tandem with Lightning’s \texttt{Trainer} to coordinate the entire training process. The Manager abstracts away many tedious engineering details—such as automatic checkpoint handling in cluster environments, consistent logging, and monitoring (optional)—so that researchers can focus on experimentation rather than infrastructure.

\paragraph{Manager and logging-everything.}
The \texttt{Manager} works synergistically with Lightning’s \texttt{Trainer} to orchestrate the entire training pipeline, handling model execution, checkpointing, and environment-specific details such as automated reloads on clusters. At the same time, it embodies our \emph{log-everything} ethos as a first-class concern: every component of the pipeline is logged in a fine-grained and structured manner. This design turns monitoring, reproducibility, and debugging into routine features rather than burdens, aligning the library’s ergonomics with the pace and reliability needs of rapid foundation model research.

\paragraph{Dictionary-first design.}
Everything in \texttt{stable-pretraining} speaks dictionaries. Datasets emit dictionary-shaped batches; modules consume and produce dictionaries; callbacks read/write named fields. Common keys include \texttt{image}, \texttt{label}, \texttt{embedding}, \texttt{loss}. This uniform interface removes glue code, keeps components swappable, and makes pipelines easy to extend.

\paragraph{Data and module composition.}
The \texttt{DataModule} encapsulates training and validation dataloaders (e.g., from Hugging Face datasets or custom sources). The \texttt{Module} bundles any number of PyTorch components (such as backbones, projectors, classifiers, or losses) and orchestrates their interaction through a user-defined \verb|forward(self, batch, stage)|. Unlike PyTorch Lightning, where one must implement separate \texttt{training\_step}, \texttt{validation\_step}, and related methods, this framework consolidates all computation in the \texttt{forward} function. The \texttt{forward} not only produces embeddings, predictions, or other intermediate representations, but can also compute losses directly when invoked during training. The return value is a dictionary that may contain arbitrary keys (e.g., ``embedding'', ``prediction'') for monitoring and analysis, with the special convention that a \texttt{``loss''} key---if present---will be used automatically for optimization. This design keeps training logic explicit and flexible while avoiding boilerplate, and it ensures that outputs, metrics, and losses are unified in a single, stage-aware interface.

\paragraph{Callbacks.}
A major convenience of our library is its set of plug-and-play callbacks for monitoring and evaluation: linear and non-linear (attentive) probes, $k$-NN probes, and collapse detection metrics (RankMe, LiDAR), among others. The callback engine is backed by an intelligent, shared-memory queue: when multiple callbacks consume the same tensors (e.g., embeddings), computations are deduplicated and memory is reused. Our callbacks deliver (i) real-time feedback on representation quality, (ii) early detection of collapse, and (iii) multi-metric views that turn debugging into insight—with minimal overhead. Importantly, all callbacks are implemented as native Lightning callbacks, ensuring full compatibility: researchers can freely mix and match our probes and monitors with any standard or custom Lightning callback in a single training loop.

\subsection{Accelerating Research}

Beyond faithfully reproducing existing approaches, \texttt{stable-pretraining} is designed to accelerate the process of exploring new ideas. Its modularity and plug-and-play utilities enable experiments that would otherwise require considerable and repetitive effort to be carried out with minimal setup. We illustrate this through two case studies. As a sanity check, we also report the linear probe accuracy over a wide range of datasets for different methods in \cref{table:probe}.

\begin{figure}[H]
    \centering
    \includegraphics[scale=0.5]{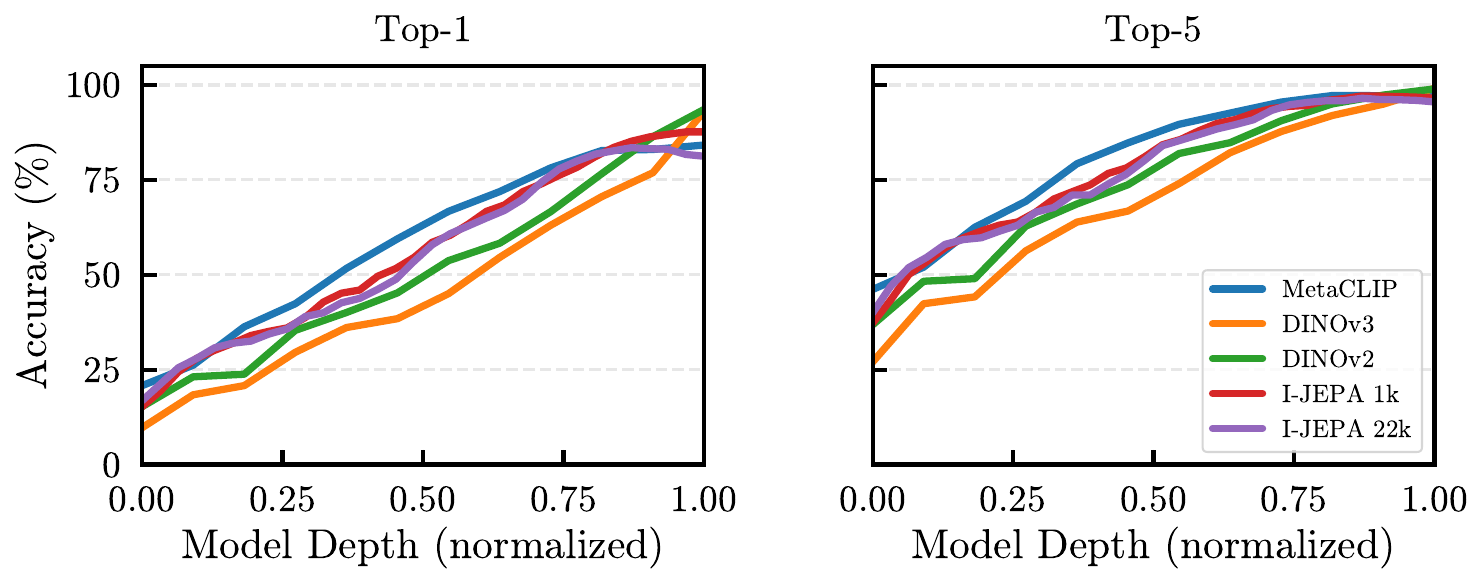}
    \caption{Depth-wise representation probing (ImageNet-100). We report the top-1 and top-5 validation accuracies of linear probes from different layers of  SOTA vision self-supervised learning methods after 100 epochs. MetaCLIP outperforms other approaches on beginning and intermediate layers, while DINOv2-3 outperforms on the last layer.}
    \label{fig:model-depth}
\end{figure}

\paragraph{Depth-wise representation probing.}  
Analyzing intermediate representations in large models typically demands intrusive modifications to training code and custom evaluation pipelines. With \texttt{stable-pretraining}, this becomes trivial: adding a linear probe at arbitrary layers requires only a few lines of configuration. As a demonstration, we probe ImageNet-100 representations at multiple depths across several state-of-the-art vision SSL models. Results (Figure~\ref{fig:model-depth}) confirm the expected trend that later layers yield stronger performance, while also revealing that MetaCLIP \citep{xu2023demystifying} excels at earlier and intermediate layers, whereas DINOv2-3 \citep{oquab2023dinov2, simeoni2025dinov3} dominates at the final layer. This experiment, often prohibitively cumbersome, is reduced to a straightforward plug-and-play setup.

\paragraph{CLIP degradation under synthetic data fine-tuning.}  
We further showcase how \texttt{stable-pretraining} facilitates rapid exploration of new research questions. Starting from a frozen CLIP ViT-B/32 \citep{radford2021learningtransferablevisualmodels} checkpoint, we continue pretraining for 8 epochs on a synthetic image dataset DiffusionDB-2M \citep{wangDiffusionDBLargescalePrompt2022}, monitoring zero-shot transfer throughout. As shown in Table~\ref{tab:clip_performance}, performance degrades sharply: Top-1 accuracy on ImageNet-100 drops by 19\% after just a single epoch, with continued training yielding no recovery. This highlights how quickly synthetic data can harm representation quality in self-supervised learning—a result that can be obtained with minimal overhead using our framework.

\begin{table}[h!]
\centering
\caption{CLIP (OpenAI clip-vit-base-patch32 d) model accuracies on ImageNet-100 validation set before and after finetuning on DiffusionDB2M. Synthetic data fine-tuning seems to degrade the quality of learned SSL representations.}
\begin{adjustbox}{max width=\textwidth}
\begin{tabular}{lccccccccc}
\toprule
Metric & No Finetuning & Epoch 1 & Epoch 2 & Epoch 3 & Epoch 4 & Epoch 5 & Epoch 6 & Epoch 7 & Epoch 8 \\
\toprule
Top-1  & \textbf{77.7} & 59.0 & 50.3 & 48.2 & 50.1 & 49.7 & 49.9 & 50.2 & 50.4 \\
Top-5  & \textbf{94.6} & 87.4 & 81.9 & 79.8 & 81.5 & 80.4 & 80.2 & 80.8 & 81.3 \\
Top-10 & \textbf{97.3} & 93.7 & 89.8 & 88.8 & 90.0 & 89.3 & 88.7 & 88.6 & 89.3 \\
\bottomrule
\end{tabular}
\end{adjustbox}
\label{tab:clip_performance}
\end{table}

\section{Conclusion}

We introduced \texttt{stable-pretraining}, an open-source library designed to accelerate and simplify research on foundation models and self-supervised learning. Built on top of PyTorch, Lightning, Hugging Face, and TorchMetrics, it ensures stability and extensibility while avoiding redundant engineering. Unlike prior efforts focused on reproducing existing methods, \texttt{stable-pretraining} emphasizes flexibility, iteration speed, and modularity by consolidating essential SSL utilities—such as probes, collapse detection, and extensible evaluation pipelines—into a unified, performance-optimized framework. A central design principle is \emph{logging everything}, making monitoring and debugging transparent, reproducible, and directly useful for research. We validate that the library not only reproduces state-of-the-art performance but also enables new research insights with minimal effort, lowering the barrier to entry while supporting large-scale experimentation.

\bibliography{ref}

@inproceedings{wolf-etal-2020-transformers,
    title = "Transformers: State-of-the-Art Natural Language Processing",
    author = "Thomas Wolf and Lysandre Debut and Victor Sanh and Julien Chaumond and Clement Delangue and Anthony Moi and Pierric Cistac and Tim Rault and Rémi Louf and Morgan Funtowicz and Joe Davison and Sam Shleifer and Patrick von Platen and Clara Ma and Yacine Jernite and Julien Plu and Canwen Xu and Teven Le Scao and Sylvain Gugger and Mariama Drame and Quentin Lhoest and Alexander M. Rush",
    booktitle = "Proceedings of the 2020 Conference on Empirical Methods in Natural Language Processing: System Demonstrations",
    month = oct,
    year = "2020",
    address = "Online",
    publisher = "Association for Computational Linguistics",
    url = "https://www.aclweb.org/anthology/2020.emnlp-demos.6",
    pages = "38--45"
}

@inproceedings{lhoest-etal-2021-datasets,
    title = "Datasets: A Community Library for Natural Language Processing",
    author = "Lhoest, Quentin  and
      Villanova del Moral, Albert  and
      Jernite, Yacine  and
      Thakur, Abhishek  and
      von Platen, Patrick  and
      Patil, Suraj  and
      Chaumond, Julien  and
      Drame, Mariama  and
      Plu, Julien  and
      Tunstall, Lewis  and
      Davison, Joe  and
      {\v{S}}a{\v{s}}ko, Mario  and
      Chhablani, Gunjan  and
      Malik, Bhavitvya  and
      Brandeis, Simon  and
      Le Scao, Teven  and
      Sanh, Victor  and
      Xu, Canwen  and
      Patry, Nicolas  and
      McMillan-Major, Angelina  and
      Schmid, Philipp  and
      Gugger, Sylvain  and
      Delangue, Cl{\'e}ment  and
      Matussi{\`e}re, Th{\'e}o  and
      Debut, Lysandre  and
      Bekman, Stas  and
      Cistac, Pierric  and
      Goehringer, Thibault  and
      Mustar, Victor  and
      Lagunas, Fran{\c{c}}ois  and
      Rush, Alexander  and
      Wolf, Thomas",
    booktitle = "Proceedings of the 2021 Conference on Empirical Methods in Natural Language Processing: System Demonstrations",
    month = nov,
    year = "2021",
    address = "Online and Punta Cana, Dominican Republic",
    publisher = "Association for Computational Linguistics",
    url = "https://aclanthology.org/2021.emnlp-demo.21",
    pages = "175--184",
    eprint={2109.02846},
    archivePrefix={arXiv},
    primaryClass={cs.CL},
}

@software{Susmelj2020lightly,
  author       = {Igor Susmelj and Matthias Heller and Philipp Wirth and Jeremy Prescott and Malte Ebner and et al.},
  title        = {Lightly},
  year         = {2020},
  url          = {https://github.com/lightly-ai/lightly},
}

@misc{goyal2021vissl,
  author =       {Priya Goyal and Quentin Duval and Jeremy Reizenstein and Matthew Leavitt and Min Xu and
                  Benjamin Lefaudeux and Mannat Singh and Vinicius Reis and Mathilde Caron and Piotr Bojanowski and
                  Armand Joulin and Ishan Misra},
  title =        {VISSL},
  howpublished = {\url{https://github.com/facebookresearch/vissl}},
  year =         {2021}
}

@article{solo-learn,
  author  = {Victor Guilherme Turrisi da Costa and Enrico Fini and Moin Nabi and Nicu Sebe and Elisa Ricci},
  title   = {solo-learn: A Library of Self-supervised Methods for Visual Representation Learning},
  journal = {Journal of Machine Learning Research},
  year    = {2022},
  volume  = {23},
  number  = {56},
  pages   = {1-6},
  url     = {http://jmlr.org/papers/v23/21-1155.html}
}

@incollection{pytorch,
title = {PyTorch: An Imperative Style, High-Performance Deep Learning Library},
author = {Paszke, Adam and Gross, Sam and Massa, Francisco and Lerer, Adam and Bradbury, James and Chanan, Gregory and Killeen, Trevor and Lin, Zeming and Gimelshein, Natalia and Antiga, Luca and Desmaison, Alban and Kopf, Andreas and Yang, Edward and DeVito, Zachary and Raison, Martin and Tejani, Alykhan and Chilamkurthy, Sasank and Steiner, Benoit and Fang, Lu and Bai, Junjie and Chintala, Soumith},
booktitle = {Advances in Neural Information Processing Systems 32},
editor = {H. Wallach and H. Larochelle and A. Beygelzimer and F. d\textquotesingle Alch\'{e}-Buc and E. Fox and R. Garnett},
pages = {8024--8035},
year = {2019},
publisher = {Curran Associates, Inc.},
url = {http://papers.neurips.cc/paper/9015-pytorch-an-imperative-style-high-performance-deep-learning-library.pdf}
}

@software{lightning,
author = {Falcon, William and {The PyTorch Lightning team}},
doi = {10.5281/zenodo.3828935},
license = {Apache-2.0},
month = mar,
title = {{PyTorch Lightning}},
url = {https://github.com/Lightning-AI/lightning},
version = {1.4},
year = {2019}
}

@inproceedings{garrido2023rankme,
  title={Rankme: Assessing the downstream performance of pretrained self-supervised representations by their rank},
  author={Garrido, Quentin and Balestriero, Randall and Najman, Laurent and Lecun, Yann},
  booktitle={International conference on machine learning},
  pages={10929--10974},
  year={2023},
  organization={PMLR}
}

@article{oquab2023dinov2,
  title={Dinov2: Learning robust visual features without supervision},
  author={Oquab, Maxime and Darcet, Timoth{\'e}e and Moutakanni, Th{\'e}o and Vo, Huy and Szafraniec, Marc and Khalidov, Vasil and Fernandez, Pierre and Haziza, Daniel and Massa, Francisco and El-Nouby, Alaaeldin and others},
  journal={arXiv preprint arXiv:2304.07193},
  year={2023}
}

@article{thilak2023lidar,
  title={Lidar: Sensing linear probing performance in joint embedding ssl architectures},
  author={Thilak, Vimal and Huang, Chen and Saremi, Omid and Dinh, Laurent and Goh, Hanlin and Nakkiran, Preetum and Susskind, Joshua M and Littwin, Etai},
  journal={arXiv preprint arXiv:2312.04000},
  year={2023}
}

@article{xu2023demystifying,
  title={Demystifying clip data},
  author={Xu, Hu and Xie, Saining and Tan, Xiaoqing Ellen and Huang, Po-Yao and Howes, Russell and Sharma, Vasu and Li, Shang-Wen and Ghosh, Gargi and Zettlemoyer, Luke and Feichtenhofer, Christoph},
  journal={arXiv preprint arXiv:2309.16671},
  year={2023}
}

@inproceedings{he2022masked,
  title={Masked autoencoders are scalable vision learners},
  author={He, Kaiming and Chen, Xinlei and Xie, Saining and Li, Yanghao and Doll{\'a}r, Piotr and Girshick, Ross},
  booktitle={Proceedings of the IEEE/CVF conference on computer vision and pattern recognition},
  pages={16000--16009},
  year={2022}
}

@article{balestriero2023cookbook,
  title={A cookbook of self-supervised learning},
  author={Balestriero, Randall and Ibrahim, Mark and Sobal, Vlad and Morcos, Ari and Shekhar, Shashank and Goldstein, Tom and Bordes, Florian and Bardes, Adrien and Mialon, Gregoire and Tian, Yuandong and others},
  journal={arXiv preprint arXiv:2304.12210},
  year={2023}
}

@article{Detlefsen2022TorchMetrics,
  title   = {TorchMetrics - Measuring Reproducibility in PyTorch},
  author  = {Detlefsen, Nicki Skafte and Borovec, Jiri and Schock, Justus and Harsh, Ananya and Koker, Teddy and Di Liello, Luca and Stancl, Daniel and Quan, Changsheng and Grechkin, Maxim and Falcon, William},
  journal = {Journal of Open Source Software},
  volume  = {7},
  number  = {70},
  pages   = {4101},
  year    = {2022},
  doi     = {10.21105/joss.04101},
  url     = {https://www.pytorchlightning.ai}
}

@article{simeoni2025dinov3,
  title={DINOv3},
  author={Sim{\'e}oni, Oriane and Vo, Huy V and Seitzer, Maximilian and Baldassarre, Federico and Oquab, Maxime and Jose, Cijo and Khalidov, Vasil and Szafraniec, Marc and Yi, Seungeun and Ramamonjisoa, Micha{\"e}l and others},
  journal={arXiv preprint arXiv:2508.10104},
  year={2025}
}

@misc{radford2021learningtransferablevisualmodels,
      title={Learning Transferable Visual Models From Natural Language Supervision}, 
      author={Alec Radford and Jong Wook Kim and Chris Hallacy and Aditya Ramesh and Gabriel Goh and Sandhini Agarwal and Girish Sastry and Amanda Askell and Pamela Mishkin and Jack Clark and Gretchen Krueger and Ilya Sutskever},
      year={2021},
      eprint={2103.00020},
      archivePrefix={arXiv},
      primaryClass={cs.CV},
      url={https://arxiv.org/abs/2103.00020}, 
}

@article{wangDiffusionDBLargescalePrompt2022,
  title = {{{DiffusionDB}}: {{A}} Large-Scale Prompt Gallery Dataset for Text-to-Image Generative Models},
  author = {Wang, Zijie J. and Montoya, Evan and Munechika, David and Yang, Haoyang and Hoover, Benjamin and Chau, Duen Horng},
  year = {2022},
  journal = {arXiv:2210.14896 [cs]},
  url = {https://arxiv.org/abs/2210.14896}
}

@inproceedings{assran2023self,
  title={Self-supervised learning from images with a joint-embedding predictive architecture},
  author={Assran, Mahmoud and Duval, Quentin and Misra, Ishan and Bojanowski, Piotr and Vincent, Pascal and Rabbat, Michael and LeCun, Yann and Ballas, Nicolas},
  booktitle={Proceedings of the IEEE/CVF Conference on Computer Vision and Pattern Recognition},
  pages={15619--15629},
  year={2023}
}

@inproceedings{caron2021emerging,
  title={Emerging properties in self-supervised vision transformers},
  author={Caron, Mathilde and Touvron, Hugo and Misra, Ishan and J{\'e}gou, Herv{\'e} and Mairal, Julien and Bojanowski, Piotr and Joulin, Armand},
  booktitle={Proceedings of the IEEE/CVF international conference on computer vision},
  pages={9650--9660},
  year={2021}
}

\appendix

\section{Code snippets.}

\begin{lstlisting}[style=sslpython, caption={SimCLR training on CIFAR-10 with \texttt{stable\_pretraining} and PyTorch Lightning}, label={lst:simclr_cifar}]
import lightning as pl
import torch
import torchmetrics
import torchvision
from torch import nn
from lightning.pytorch.loggers import WandbLogger

import stable_pretraining as spt
from stable_pretraining.data import transforms

# Define augmentations for SimCLR (creates 2 views of each image)
simclr_transform = transforms.MultiViewTransform(
    [
        transforms.Compose(
            transforms.RGB(),
            transforms.RandomResizedCrop((32, 32), scale=(0.2, 1.0)),
            transforms.RandomHorizontalFlip(p=0.5),
            transforms.ColorJitter(brightness=0.4, contrast=0.4, saturation=0.2, hue=0.1, p=0.8),
            transforms.RandomGrayscale(p=0.2),
            transforms.ToImage(**spt.data.static.CIFAR10),
        ),
        # Second view with slightly different augmentations
        transforms.Compose(
            transforms.RGB(),
            transforms.RandomResizedCrop((32, 32), scale=(0.08, 1.0)),
            transforms.RandomHorizontalFlip(p=0.5),
            transforms.ColorJitter(brightness=0.4, contrast=0.4, saturation=0.2, hue=0.1, p=0.8),
            transforms.RandomGrayscale(p=0.2),
            transforms.RandomSolarize(threshold=0.5, p=0.2),
            transforms.ToImage(**spt.data.static.CIFAR10),
        ),
    ]
)

# Load CIFAR-10 and wrap in dictionary format
cifar_train = torchvision.datasets.CIFAR10(train=True)
cifar_val = torchvision.datasets.CIFAR10(train=False)

train_dataset = spt.data.FromTorchDataset(
    cifar_train,
    names=["image", "label"],  # Convert tuple to dictionary
    transform=simclr_transform,
)

val_dataset = spt.data.FromTorchDataset(
    cifar_val,
    names=["image", "label"],
    transform=transforms.Compose(
        transforms.RGB(),
        transforms.Resize((32, 32)),
        transforms.ToImage(**spt.data.static.CIFAR10),
    ),
)

# Create dataloaders with view sampling for contrastive learning
train_dataloader = torch.utils.data.DataLoader(
    dataset=train_dataset,
    sampler=spt.data.sampler.RepeatedRandomSampler(train_dataset, n_views=2),
    batch_size=256,
    num_workers=8,
    drop_last=True,
)

val_dataloader = torch.utils.data.DataLoader(
    dataset=val_dataset,
    batch_size=256,
    num_workers=10,
)

data = spt.data.DataModule(train=train_dataloader, val=val_dataloader)

# Define the forward function (replaces training_step in PyTorch Lightning)
def forward(self, batch, stage):
    out = {}
    out["embedding"] = self.backbone(batch["image"])
    if self.training:
        # Project embeddings and compute contrastive loss
        proj = self.projector(out["embedding"])
        views = spt.data.fold_views(proj, batch["sample_idx"])
        out["loss"] = self.simclr_loss(views[0], views[1])
    return out

# Build model components
backbone = spt.backbone.from_torchvision("resnet18", low_resolution=True)
backbone.fc = torch.nn.Identity()  # Remove classification head

projector = nn.Sequential(
    nn.Linear(512, 2048),
    nn.BatchNorm1d(2048),
    nn.ReLU(inplace=True),
    nn.Linear(2048, 2048),
    nn.BatchNorm1d(2048),
    nn.ReLU(inplace=True),
    nn.Linear(2048, 256),
)

# Create the module with all components
module = spt.Module(
    backbone=backbone,
    projector=projector,
    forward=forward,
    simclr_loss=spt.losses.NTXEntLoss(temperature=0.5),
    optim={
        "optimizer": {"type": "LARS", "lr": 5, "weight_decay": 1e-6},
        "scheduler": {"type": "LinearWarmupCosineAnnealing"},
        "interval": "epoch",
    },
)

# Add callbacks for monitoring performance during training
linear_probe = spt.callbacks.OnlineProbe(
    name="linear_probe",
    input="embedding",
    target="label",
    probe=torch.nn.Linear(512, 10),
    loss_fn=torch.nn.CrossEntropyLoss(),
    metrics={
        "top1": torchmetrics.classification.MulticlassAccuracy(10),
        "top5": torchmetrics.classification.MulticlassAccuracy(10, top_k=5),
    },
)

knn_probe = spt.callbacks.OnlineKNN(
    name="knn_probe",
    input="embedding",
    target="label",
    queue_length=20000,
    metrics={"accuracy": torchmetrics.classification.MulticlassAccuracy(10)},
    input_dim=512,
    k=10,
)

# Configure training
trainer = pl.Trainer(
    max_epochs=1000,
    callbacks=[knn_probe, linear_probe],  # Monitor SSL quality in real-time
    precision="16-mixed",
    logger=WandbLogger(project="cifar10-simclr"),
)

# Launch training
manager = spt.Manager(trainer=trainer, module=module, data=data)
manager()
\end{lstlisting}

\end{document}